\documentclass[prb,twocolumn,showpacs,amsmath,amssymb]{revtex4}

\usepackage{amsmath}
\usepackage{graphicx}
\usepackage[pdftex,dvips]{epsfig}
\newcommand{\be}{\begin{equation}}
\newcommand{\ee}{\end{equation}}
\newcommand{\bma}{\begin{displaymath}}
\newcommand{\ema}{\end{displaymath}}
 
\begin{document}

\title{Magnetic phases of one-dimensional lattices with 2 to 4 fermions
per site}

\author{J.-P. Nikkarila$^{1,2}$, M. Koskinen$^1$, S.M. Reimann$^3$ and M. Manninen$^1$}

\affiliation{\sl $^1$NanoScience Center, Department of Physics,
FIN-40014 University of Jyv\"askyl\"a, Finland}

\affiliation{\sl $^2$Inspecta LTD, FI-02151 Espoo, Finland}

\affiliation{\sl $^3$Mathematical Physics, LTH, Lund University, 
SE-22100 Lund, Sweden}
 
\date{\today}

\begin{abstract} 
We study the spectral and magnetic properties of one-dimensional lattices 
filled with 2 to 4 fermions (with spin 1/2) per lattice site.
We use a generalized Hubbard model that takes account all interactions 
on a lattice site, and solve the many-particle problem by exact
diagonalization. We find an intriguing magnetic phase diagram 
which includes ferromagnetism, spin-one Heisenberg antiferromagnetism,
and orbital antiferromagnetism. 
\end{abstract}
%\pacs{71.10.-w, 73.21.La,03.75.Hh,03.75.Ss}

\maketitle

\section{introduction}

Artificial lattices resemble periodic arrangements of quantum wells
confining a small number of particles. Experimentally, both lateral and
vertical lattice structures can be realized. 
Examples are arrays of quantum dots in semiconductor heterostructures 
\cite{lee2001,schmidbauer2006,kohmoto2002} confining the conduction electrons, 
or optical lattices -- stable periodic arrays of potentials created by
standing waves of laser light~\cite{eurphysnews,jaksch2005}.
Varying the intensity of the
laser light, one can change the depths  of the single traps, i.e. the single
sites.  In such egg-box like potentials, experimentalists 
can confine ultra-cold atoms, of bosonic or fermionic %%@
character~\cite{greiner2002,han2000,kerman2000,modugno2003,rom2006,chin2006},
achieving particle numbers on the sites that are even less than 
three.
The strengths and even the sign of the interactions between the atoms can 
be tuned by Feshbach %%@
resonances~\cite{feshbach1958,inouye1998,courteille1998,roberts1998,duine2004,theis2004}. 

The basic difference between artificial lattices and normal lattices 
(such as the crystal structure of solids) is, that in artificial
lattices the particles confined in the lattice do not 
play any role for determining the intrinsic lattice structure. 
A possible degeneracy of the many-particle states can then not be
removed by lattice distortion. Instead, it may lead to 
internal symmetry breaking and, for example, to spontaneous magnetism
and superconductivity.
Recent experiments have inspired much theoretical
work on artificial lattices, both with cold 
atoms~\cite{gu2007,xianlong2007,massel2005,koponen2006,koponen2007} and
quantum dots~\cite{chen1997,taut2000}.

Mean-field calculations based on the spin-density functional
theory predict that Hund's first
rule determines the total spin of an isolated, individual lattice 
site~\cite{koskinen1997,reimann2002}. The magnetism of the 
lattice as a whole then depends on the total spin of the individual
lattice sites, on the lattice structure and 
on the coupling between the 
sites\cite{koskinen2003,karkkainen2007,karkkainen2007a,karkkainen2007b}.
A simple tight-binding model with a few parameters can account for most of the 
these findings\cite{koskinen2003b}.
Related results have been obtained for quantum dot 
molecules using the density functional method\cite{kolehmainen2000}.

The eigenstates of single quantum dots with a 
few electrons can be calculated ``exactly''
(i.e. to a high degree of convergence with respect to the 
necessary restrictions
in Hilbert space) by diagonalizing the many-body Hamiltonian 
(for a review see Ref.\cite{reimann2002}).
Methods beyond the mean-field approximation have also been applied 
to quantum dot %%@
molecules\cite{yannouleas1999,bayer2001,harju2002,mireles2006,scheibner2007,zhang2007}.

For a lattice with strongly correlated particles the generic model is
the Hubbard model, which has been amply studied in the case of one state
per lattice site (for reviews see\cite{voit1994,kolomeisky1996}).
From an experimental viewpoint, it has been argued that the Hubbard 
approach is ideal for describing contact-interacting atoms 
in an optical lattice~\cite{jaksch98,greiner02,stoferle2004,xu2005,jaksch2005}.
The one-dimensional Hubbard model is exactly solvable using the
Bethe ansatz\cite{lieb1968}.
The magnetism of finite molecules\cite{pastor1994,lopezurias1999}
and quantum rings\cite{viefers2004} have also been studied in the 
simple Hubbard model.

The purpose of this paper is to study the magnetism of an
artificial one-dimensional (1D) lattice in the case 
where the lattice site is filled on average 
with 2 to 4 fermions, which can be electrons in a quantum dot
lattice or fermionic atoms in an optical lattice.
We call the electrons or atoms generally as particles. 
We assume the confining potential in each lattice site to be 
quasi-two-dimensional and nearly harmonic at the bottom.
In this case the $1s$-level of each
lattice site is filled, and the degenerate $1p$ level is partially filled.
We use a generalization of the Hubbard model to describe the
interactions: The particles interact only within a lattice site.
We solve the Hubbard Hamiltonian by exact diagonalization for a  
finite length of the lattice using periodic boundary conditions.
The results show many different magnetic structures which are
analyzed through their relations to the Heisenberg model and the simple
single-state Hubbard model.

\section{Theoretical models}

\subsection{1D lattice with $p$-orbitals}

\begin{figure}[!h]
\includegraphics[width=0.98\columnwidth]{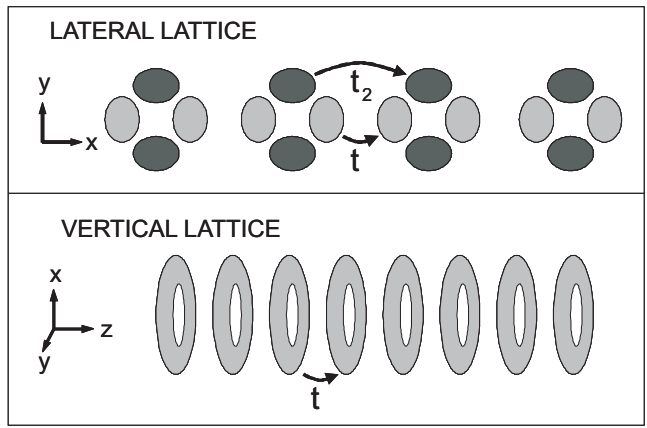} 
\caption{Schematic pictures of 1D lattices considered.
Each lattice site has $p_x$ and $p_y$ orbitals. In the {\it lateral}
case these are shown as light and
dark-gray densities. Here, the hopping probabilities $t$ and $t_2$
between neighboring lattice sites are different for $p_x$ and $p_y$ orbitals.
In the {\it vertical} case it is natural to use states with 'rotating orbitals'
$p_{+1}$ and $p_{-1}$ with circularly symmetric densities. In this case 
there is only one hopping probability $t$.
 }
\label{lattice}
\end{figure}

We consider an artificial lattice where the confining potential 
at each lattice site is nearly harmonic and quasi-two-dimensional so that
the single-particle level structure in each lattice site is 
$1s$, $1p$, $2s1d$ {\it etc.} 
We assume that in all cases the $1s$-state is 
filled completely, and the doubly degenerate $1p$-level is partially filled. 
Furthermore, we assume that the shells are well separated in energy, 
so that we can neglect the mixing of the $1p$ shell with
the $1s$ or $2s1d$ shells. This leads to a
generalized Hubbard model which has in each lattice site
only two orbitals which we call either $p_x$ and $p_y$
or $p_{-1}$ and $p_{+1}$, respectively. The latter
notation refers to orbitals with angular momentum quantum 
numbers $-1$ and $+1$
(clockwise or counterclockwise rotation of the $p$-state).

The two kinds of 1D lattices considered are  
schematically shown in Fig. \ref{lattice}.
In the case of  
semiconductor quantum dots, these are often called {\it lateral} and 
{\it vertical} structures. In the lateral lattice the hoppings
between neighboring $p_x$ and $p_y$ states are 
different and denoted by $t$ and $t_2$, where $t_2<t$.
%We call these two $p$-states also as 'longitudinal' and
%'perpendicular' state
For the vertical lattice,
it is natural to use the angular momentum states
$p_{-1}$ and $p_{+1}$.  In that case there is only one
hopping parameter $t$ (or equivalently $t_2=t$). Note that for the
single-particle wave functions we have $\psi_{+1}=(\psi_{p_x}+i\psi_{p_y})/\sqrt{2}$ 
and $\psi_{-1}=(\psi_{p_x}-i\psi_{p_y})/\sqrt{2}$.

\subsection{Hubbard model}

We assume a generalized Hubbard model Hamiltonian
\be
\hat{H}=\hat{J}+\hat{U},
\label{hamiltonian}
\ee
where the first term represents inter-site hoppings between neighboring 
lattice sites and the second term intra-site two-body interactions. 

Hoppings preserve spin, and are equal for spin-up and spin-down 
particles. Thus, $\hat J$ separates into two symmetric spin parts:
$\hat J = \sum _{\sigma = \uparrow, \downarrow} \hat J _{\sigma }$.
For our one-dimensional lattice with $p$-orbitals
\begin{equation}
\hat {J}_{\sigma } = - \sum _n \sum _{jj'} J_{jj'}\left( 
c^{\dagger } _{nj\sigma } c _{n+1 j' \sigma } + h.c. \right),
\label{hopping}
\end{equation}
where $n$ is the lattice site index, and $j$ and $j'$ denote the 
$p$-orbital in question. (In the simple Hubbard model, there 
would be only one space state per site, and the $j$-indices not needed). 

Some of the hopping integrals $J_{jj'}$ are zero due to symmetry. 
The non-zero integrals are treated as essentially free model parameters, 
$t$ and $t_2$. Thus, we have 
\bma
\begin{array}{lll}
j & j' & J_{jj'} \\
\hline
p_x & p_x & t \\
p_y & p_y & t_2 \\
p_x & p_y & 0 \\
\hline
\end{array}\hskip1cm
\begin{array}{lll}
j & j' & J_{jj'} \\
\hline
p_{-1} & p_{-1} & t \\
p_{+1} & p_{+1} & t \\
p_{-1} & p_{+1} & 0 \\
\hline
\end{array}
\ema
for the lateral and vertical lattice, respectively. 
(Note, that in the case $t_2=t$, the lateral model actually is identical to
the vertical model, irrespective of the different $p$-orbit basis used). 

We approximate the two-body interactions in the spirit of the  
tight-binding model:
The particles only interact when they are at the same lattice site.
Thus, $\hat U$ separates in the symmetric parts representing interactions on
each site $n$: $\hat U = \sum _n \hat U _n$.
Within a site, full (spin-indpendent) two-body interaction is allowed, 
which yields
\begin{equation}
\hat U _n = {1\over 2} \sum _{\stackrel{j_1j_2j_3j_4}{\sigma \sigma '}}
U_{j_1j_2j_3j_4} 
c_{nj_1\sigma }^{\dagger }
 c_{nj_2\sigma ' }^{\dagger }
 c_{nj_4\sigma ' }
 c_{nj_3\sigma  }
\end{equation}
where $U_{j_1j_2j_3j_4}$ are the direct space matrix elements 
of on-site interaction, depending on the interaction itself
and the $j$-orbits in question, i.e. the eigenstates of the confining 
potential. 

For contact interactions, the ratios of the different
matrix elements are independent of the confining potential, 
as long as it has circular symmetry. 
For the non-zero matrix elements (together with those obtained by 
allowed $j$-index permutations), we obtain
\bma
\begin{array}{ccccc}
j_1 & j_2 & j_3 & j_4 & U_{j_1j_2j_3j_4} \\
\hline
p_x & p_x & p_x & p_x & 3U \\
p_x & p_x & p_y & p_y & U \\
p_x & p_y & p_x & p_y & U \\
p_x & p_y & p_y & p_x & U \\
p_y & p_y & p_y & p_y & 3U \\
\hline
p_{-1} & p_{-1} & p_{-1} & p_{-1} & 2U \\
p_{-1} & p_{+1} & p_{-1} & p_{+1} & 2U \\
p_{-1} & p_{+1} & p_{+1} & p_{-1} &  2U-\Delta\\
p_{+1} & p_{+1} & p_{+1} & p_{+1} & 2U \\
\hline
\end{array}
\ema
where $U$ is the only parameter describing the strength of the interaction.
All together, we thus have three parameters $t$, $t_2$ and $U$. 
One of them can be fixed to set the energy scale.
We choose this to be $t$ and represent the results for $t=1$
(all energies are given in units of $t$).
In some cases with vertical lattices we also consider 
an interaction of finite width. 
This can be mimicked by decreasing one of the matrix elements by a small
amount $\Delta$, as indicated in the above table. For contact 
interactions, $\Delta=0$.

We solve the Hamiltonian for a lattice with $L$ lattice sites 
using periodic boundary conditions ($\hat J$ connects also the 
last and the first site). The Lanczos method is used to
find the low-energy eigenvalues and eigenvectors of the Hamiltonian matrix. 
We take advantage of the periodicity of the lattice and solve
the eigenvalues separately for each Bloch $k$-value.
In practice this means that, instead of using ``site-states'' 
$\mid n j\sigma \rangle $ as a single-particle basis to span the Fock space, 
one uses Bloch states of the tight-binding model (eigenstates of $\hat J$).
In this study, the hopping does not mix the $p_x$ and $p_y$ orbitals 
in the lateral case, nor $p_{-1}$ and $p_{+1}$ orbitals
in the vertical case. We then have separate, simple bands 
with energy eigenvalues
\be
\epsilon_t(k)=-2t\cos\left(\frac{2\pi k}{L}\right),
\label{bloch}
\ee
where $k$ takes integer values $0,~1,~\cdots,L-1$. 
Note that in the lateral case, the $p_x$ and $p_y$ bands have different
widths for $t $ and $t_2$, respectively. In the vertical case, 
the widths are always the same. 

We do not take advantage of the fact that the Hamiltonian
does not depend on spin, but diagonalize the system for $S_z=0$
and only afterwards determine the total spin $S$ for each 
many-particle state.
The total number of particles is denoted by $N$ and the 
numbers of spin-up particles
and spin-down particles by $N_{\uparrow}$ and $N_{\downarrow}$.
We note that because of the spin degree of freedom, 
the maximum number of particles in a lattice with length $L$,
is $N_{\text{max}}=4L$. The filling fraction
by $\nu=N/L$ gets values from 0 to 4. In this study we consider
only the region $\nu=0\cdots 2$. Due to the symmetry of the Hamiltonian
the region $\nu=2\cdots 4$ will have similar properties.

As discussed earlier, we assume that the hopping can occur only between 
the nearest neighbours. 
It should be noted, however, that the interaction part
of the Hamiltonian allows {\it intra-site hopping}, via 
scattering from one single-particle state to another inside any
lattice site. This becomes important especially in the case where $t_2=0$,
where the hopping only occurs through the $p_x$ states. 

\subsection{Heisenberg model}

It is well-known that the simple Hubbard model in the limit of large $U/t$ 
approaches the antiferromagnetic Heisenberg model. 
In this case, the low-lying eigenstates are characterized by one spin $1/2$
particle on each site. In a similar way, in some limiting cases, our results 
with $p$-orbitals approach those of the Heisenberg model with $S=1$ (two
particles on each site with aligned spins), or with $S=1/2 $ (polarized 
system with one
fermion on each site, with the $p$-orbitals playing the role of the 
spin components). 
The effective Hamiltonian is then
\be
\hat{H}_{\rm eff}=J_{\rm eff}\sum {\bf S}_n \cdot {\bf S}_{n+1} +\text{constant}
\label{heisenberg},
\ee
where $J_{\rm eff}$ is the effective exchange interaction and 
${\bf S}_n$ the spin operator for site $n$. 
We compare the Heisenberg and Hubbard model 
for the case of four sites, $L=4$, where the spectrum of the antiferromagnetic 
Heisenberg model is exactly solvable\cite{ashcroft1976,viefers2004}.

\section{Results}

\subsection{A single lattice site with two particles}

A single site with two particles obeys Hund's first rule
to maximize the spin. The energy difference between the lowest
$S=0$ and $S=1$ states is the 'exchange splitting' and equals 
$\Delta E = E_{S=0} - E_{S=1}=2U$. In the case of a finite-range 
interaction the exchange splitting is $2U-\Delta$.
Table \ref{singleatom} gives the 
energy spectrum of a single lattice site.
We will see below that in the limit of large $U$ the half-filled system
($\nu=N/L=2$) becomes a Heisenberg antiferromagnet with $S=1$.

\begin{table}[!h]
\caption{Energy levels and corresponding total spin of a single lattice site
with two particles.}
\begin{tabular}{ccc}
\hline
State No & $E$ & $S$ \\
\hline
1 & $\Delta$ & 1 \\
2 & $2U$ & 0 \\
3 & $2U$ & 0 \\
4 & $4U-\Delta$ & 0 \\
\hline
\label{singleatom}
\end{tabular}
\end{table}

\subsection{Half-filled vertical lattice: $N=2L$}

\begin{figure}[!h]
\includegraphics[angle=-90,width=0.9\columnwidth]{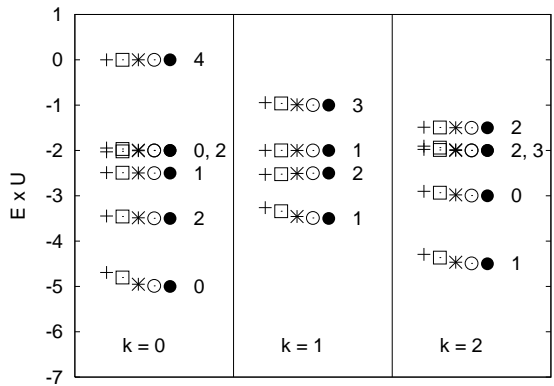} 
%\vspace{0.5cm}   
\caption{Low-energy levels ($E\ll U$) for $L=4$ and $N=8$
calculated for different values of $U$. The numbers next to the
levels denote the total
spin $S$ of the many-particle state. 
The wave vector $k$ has values 0, 1, and 2.
The symbols plus, square, star, circle and dot correspond to
$U$ values 2, 2.5, 5, 10, and 50, respectively. The energy levels
for $U=50$ agree with those of the $S=1$ Heisenberg model with
0.01 \% accuracy.}
\label{anti4}
\end{figure}

In the half-filled case there is one particle per
orbital. When $U$ is large, each lattice site will have spin $S=1$ due
to the large exchange splitting. The only way to allow particles to
hop from one site to the neighboring one is to orient the 
total spins of neighboring sites opposite, {\it i.e.} with 
antiferromagnetic order. For ferromagnetic order, the hopping would 
be prohibited by the Pauli exclusion principle. In this case the 
total energy of the system would be zero (assuming $\Delta=0$).
For antiferromagnetic order the allowed hopping can reduce 
the energy to a slightly negative value. 

We will first study the case of vertical lattice ($t_2=t$).
Figure \ref{anti4} shows the low-energy 
energy levels for four lattice sites and eight particles ($L=4$, $N=8$),
calculated for different values of $U$. All the levels with energy 
$E\stackrel{<}{\sim } U$ are shown. For the largest value $U=50$ the 
spectrum agrees with that of the Heisenberg model 
($J_{\rm eff}=2/U$) with 0.01 \% accuracy.
The Heisenberg model for four sites is an
exacly solvable textbook problem\cite{ashcroft1976,viefers2004}.
It is interesting to notice that even for $U=2$ the spectrum is 
qualitatively still the same. Only when $U\lesssim 1.5$ 
new states start to appear in the low-energy spectrum.

\subsection{Vertical lattice polarized fermions: The noninteracting case}

We will now consider polarized fermions (e.g. electrons or 
fermionic atoms with $N=N_\uparrow$). 
For contact interactions between the fermions, the problem becomes 
non-interacting since the Pauli exclusion principle forbids
two fermions to be at the same state. The energy spectrum can then be
constructed by filling particles to the Bloch states 
(Eq. (\ref{bloch})) which are solutions
of the noninteracting Hamiltonian $\hat{J}$.

In this (trivial) case, it is important to note that 
$(i)$ each single-particle Bloch
state is doubly degenerate due to the two states per site, and $(ii)$ only for 
particle numbers $N=4n+2$ the ground state is 
non-degenerate ($n$ is a non-negative integer).
$(ii)$ implies that the ground state energy (of polarized fermions)
as a function of $N$ has local minima for $N=2,~6,~10, \cdots$.
We will see later that these special values form single domain
ferromagnets when $N<2L$.

\subsection{Orbital antiferromagnet of polarized fermions: Vertical lattice with $N=L$}

\begin{figure}[!h]
\includegraphics[angle=-90,width=0.98\columnwidth]{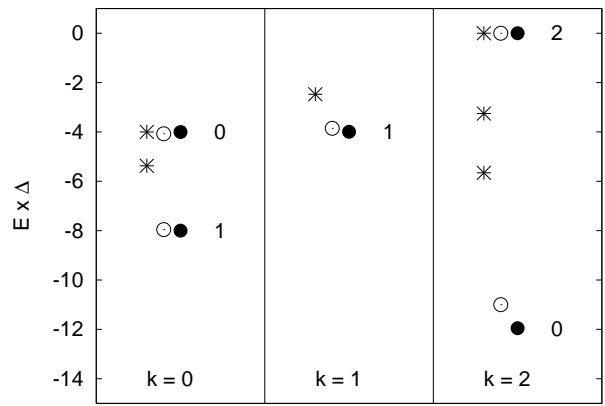} 
%\vspace{0.5cm}   
\caption{Low-energy levels ($E\le 0$) for polarized fermions 
with $L=4$ and $N=4$
calculated for different values of $\Delta$ for $U=50$. The numbers
next to the levels denote the total spin of the corresponding state of the
spin-1/2 Heisenberg model.  
The wave vector $k$ has values 0, 1, and 2.
The symbols star, circle and dot correspond to
$\Delta$ values 2, 10, and 50, respectively. The energy levels
for $\Delta=50$ agrees with those of the $S=1/2$ Heisenberg model with
0.1 \% accuracy.}
\label{orbanti4}
\end{figure}

Let us now consider polarized fermions {\it with a finite range interaction} and
one fermion per lattice site ($N=N_\uparrow=L$). 
The finite range here means only that $\Delta>0$.
However, the finite range does not lead to interaction of 
particles sitting at different lattice sites.
Each site still has two $p$ states.
The large $U$ limit in this case is an antiferromagnet where the 
'magnetic moment' in each lattice site is not the spin but the orbital angular 
momentum of the $p$ states, which can have the two values +1 or -1. 
 
There are two reasons for this state to become the ground state.
First, it costs energy (by $\Delta$) for two particles 
to occupy the same site. Thus, the particles prefer to be at different sites.
Second, the particles can only hop to the neighboring site if
they are at different orbital states. Although the particles
prefer to be at different sites, a small amount of 'virtual' 
hopping is necessary to reduce the energy.

Again,  we compare the spectrum with the exact result 
of the Heisenberg model for four particles. 
In Fig. \ref{orbanti4}, all the low energy states are plotted 
for different $k$ values. For large $U$ and $\Delta$ the agreement
between the Hubbard model and Heisenberg model becomes perfect with 
$J_{\rm eff}=1/\Delta$.

\subsection{Vertical lattices with $N<2L$: Ferromagnetism}

Next, we consider vertical lattices with contact interactions and
large values of $U$ ($U\ge 10$).
The results show that the ground states for $N=2,~6,$ and $10$ 
have maximum possible total
spin, i.e. they are ferromagnetic. This is true 
for all values of $L>N/2$ where
the computations could be performed (the matrix sizes increase very
fast with $L$). The ferromagnetic ground state can be understood as follows.
When $L\gg N/2$ the ferromagnetic state allows particles to move freely
in the lattice, as even in cases where two particles are in the same
lattice site, 
they do not interact. In other words, particles with the same spin
can pass each other without any cost of energy. 
If the particles have opposite spin, however, they suffer repulsive
interaction whenever they are at the same lattice site --
even if they are at different $p$-states. 

\begin{figure}[!h]
\includegraphics[width=0.9\columnwidth]{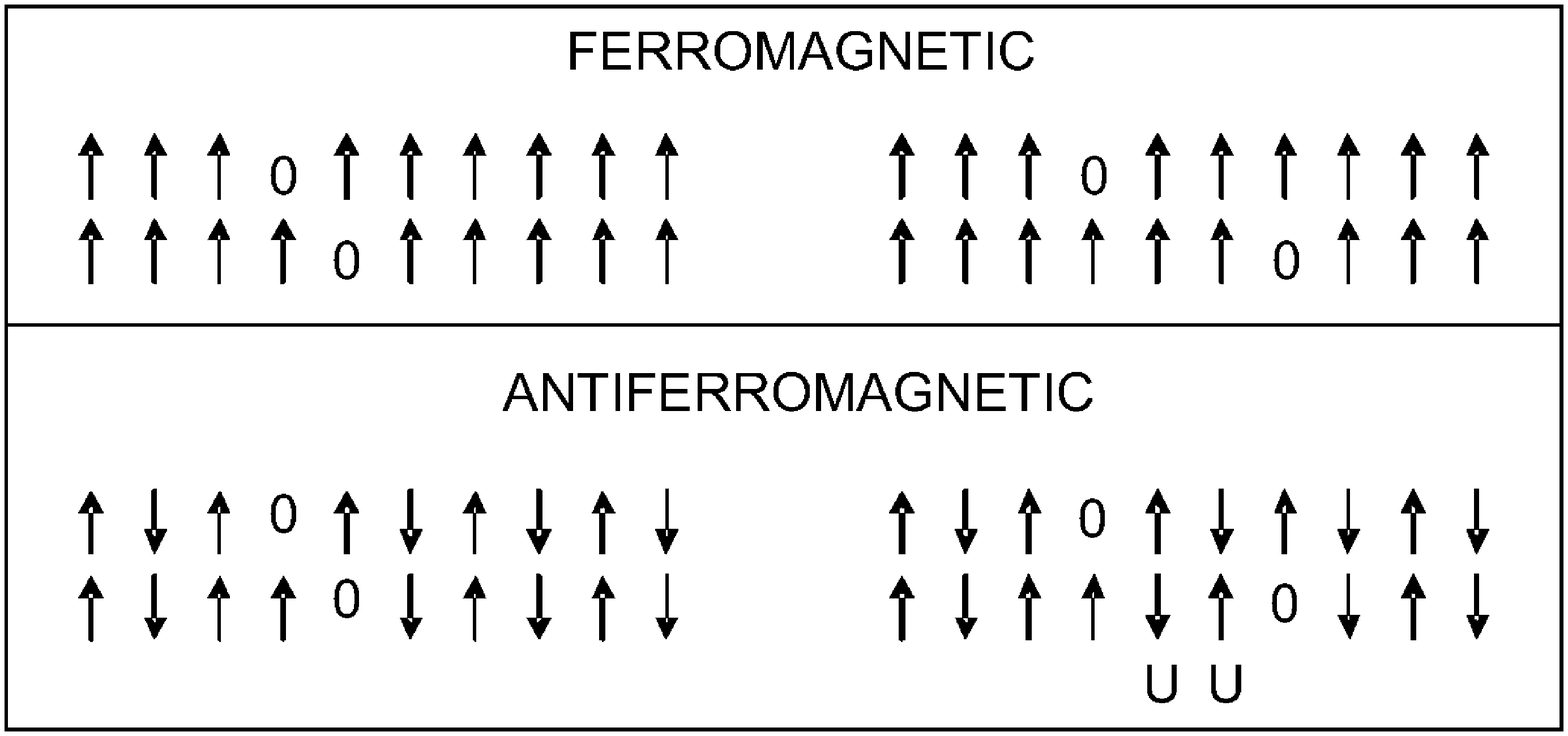} 
%\vspace{0.5cm}   
\caption{Motion of holes in the nearly half-filled case with large $U$
which prevents two opposite spins to be at the same lattice site.
In the ferromagnetic case, the holes can move independently, while in
the antiferromagnetic case they are bound together since their separation
costs energy, as indicated by $U$.}
\label{holesaf}
\end{figure}

Let us now consider what happens if we start from the antiferromagnetic 
$L=N/2$ state and increase $L$ by one. Also in this case the results show that,
independent of the particle number (here, $2\le N\le 10$), the ground state is 
ferromagnetic. In the antiferromagnetic case with $L=N/2$ the lowest
energy is proportional to $-1/U$, which for large $U$ is very small.
The energy of the ferromagnetic case with $L=N/2$ is zero since there is no
room for hopping (the single-particle bands are filled). 
If now one lattice site is added, the ferromagnetic energy
becomes $-4t$. This is because there are now two freely moving 
holes in the system, as illustrated in Fig. \ref{holesaf}.
The situation is different if the system remains antiferromagnetic.
Also in this case there are two holes, but now they are bound together,
since their separation costs energy, as illustrated in Fig. \ref{holesaf}.
The total energy of the antiferromagnetic state will necessary be above the 
ferromagnetic energy $-4t$. Consequently, adding one
lattice 
site to the antiferromagnetic $L=N/2$ case transforms it to a ferromagnetic
state. Alternatively, we can start from the half-filled case and remove
one particle. In the ferromagnetic case the hole is free and has an
energy of $-2t$, while in the antiferromagnetic case the hole is localized and 
its energy is zero.

\begin{figure}[!h]
\includegraphics[width=0.98\columnwidth]{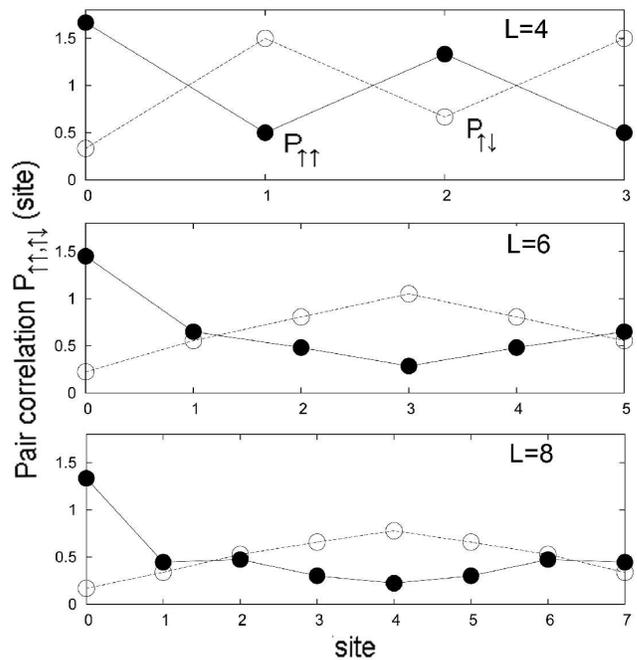} 
\caption{Pair-correlation of vertical lattices of different lengths $L=4,~6,~8$ 
with $N=8$ fermions. In each case the reference site is 0 where one particle
with spin-up is fixed. The solid line with black dots shows the total up-up
correlation and the dashed line with open circles the up-down correlation.
The uppermost panel shows the antiferromagnetic ground state of $L=4$,
the center panel and the lowest panel show the two domains of the ferromagnetic
ground states of $L=6$ and $L=8$, respectively.}
\label{afornot}
\end{figure}

As mentioned above, the ferromagnetic ground state has total spin 
$S=N/2$ for $N=2,~6,~10,~\cdots$. However, the situation is more complicated
for $N=4,~8,~12,~\cdots$. In these cases the total spin of the ground state
is $S=0$ for all $L>N/2$. Nevertheless, we argue that also these cases
are ferromagnetic, but now the ground state has a spin wave which rotates the
spin once within the length $L$. Alternatively, we can 
apply the picture that the
ferromagnetic ground state consists of two domains with 
opposite spin directions. The reason for this behavior 
is the fact that for these particle
numbers the ferromagnetic state is degenerate
and the spin wave (or domain formation) provides a way to 
remove the degeneracy and reduce the total energy.

Figure \ref{afornot} shows the pair-correlation function of
$N=8$ particles for $L=4$, 6, and 8. We fix one particle in a state,
say $p_{+1}$ with spin-up in lattice site 0 and determine the 
conditional propability of finding the other
spin-up and spin-down particles on the other lattice sites. 
Figure \ref{afornot} shows clearly that for $L=4=N/2$ the result is antiferromagnetic,
while for $L=6$ and $L=8$ the spin changes direction only once within the
length $L$,  as it would happen for the longest possible spin wave.
It is interesting to note that in fact, the system with 
two states per site is very different from that with only one
$s$ state per site. In the latter case the system remains antiferromagnetic
(for large $U$) for all values of $L$\cite{yu1992,viefers2004}.

\subsection{Lateral lattices: $t_2<t$}

In the lateral lattice,
as shown in Fig. \ref{lattice}, the hopping parameters $t$ and $t_2$
for the two $p$-states are different. 
The structure of the ground state and the many-particle spectrum 
then depends on the ratio $t_2/t$.
We will now study the magnetism as a function of this ratio
and of the filling fraction $N/L$.

For different values of $t$ and $t_2$,
for noninteracting particles we have two cosine bands,
Eq. (\ref{bloch}), which reach from
$-2t$ to $+2t$ and from $-2t_2$ to $+2t_2$, respectively.
Lets consider first the ferromagnetic case with low filling
and remember that for contact interactions the system 
becomes non-interacting. In the limit of low filling
and $t_2<t$, only the $t$-band is occupied.
This is equivalent to the simple
one-state Hubbard model. But we know that the ground 
state of the one-state Hubbard model is antiferromagnetic 
in the case of low filling. Consequently, the ground state
will be antiferromagnetic whenever the corresponding
ferromagnetic state would only
occupy the $t$-band. This condition can be easily derived, 
\be
\frac{N}{L}<\frac{1}{2\pi}\cos^{-1}(t_2/t)~.
\label{limit1}
\ee
A similar argument can be used to show that for
\be
N/L>2-\frac{1}{2\pi}\cos^{-1}(t_2/t)
\label{limit2}
\ee
the system is also antiferromagnetic. Here, the
holes in the ferromagnetic case only occupy the $t$-band.
In between these limits, both bands are partially 
filled and determining the magnetism is more complicated.
For $t_2=t$ the lateral lattice equals
the vertical lattice. We have shown above 
that this case should always be ferromagnetic. We can thus expect that
for $t_2$ close to $t$, the system is ferromagnetic.

\begin{figure}[!h]
\includegraphics[width=0.98\columnwidth]{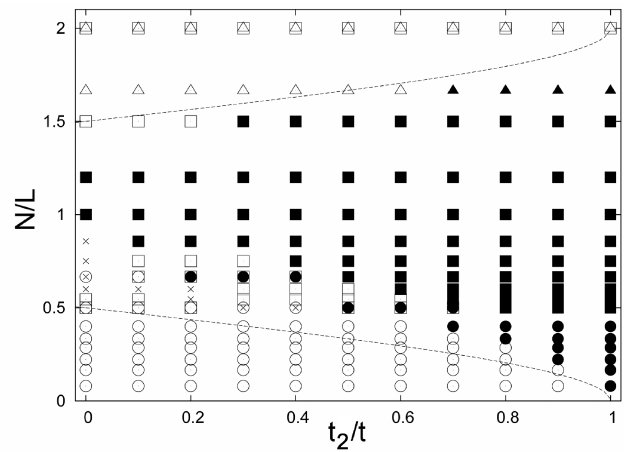}
\caption{Magnetic phase diagram of the one-dimensional lateral 
lattice. The vertical axis shows the 
number of particles per site (on $p$-states) and the lateral  
axis the ratio $t_2/t$ of the two hopping parameters.
The dashed lines show the borders between which
the narrower band ($t_2$ band) is filled. Outside this area
the lattice is antiferromagnetic (except at $t_2=t$).
The filled (open) symbols show the ferromagnetic 
(antiferromagnetic) ground states obtained with
exact diagonalization of the Hubbard Hamiltonian for 
10 (triangles), 6 (squares), and 2 (circles) particles.
Crosses show those results for $N=6$ which are not 
ferro or antiferromagnetic, i.e. $0<S<N/2$.
The numerical results are for $U=10$.
}
\label{phase1}
\end{figure}

Figure \ref{phase1} shows the magnetism of the ground state, 
as a function of both the number of particles per site (on the $p$-states) 
and the ratio $t_2/t$ of the two hopping parameters,  
calculated by diagonalizing the Hubbard
Hamiltonian. The figure also shows the limits given
by Eqs. (\ref{limit1}) and (\ref{limit2}). 
Indeed, we see that between these limits, the
ground state is mainly ferromagnetic, while outside 
these limits it is always antiferromagnetic.
Figure \ref{phase1} shows results computed for 2, 6 and 10
particles, where the ferromagnetic phase is simple
and seen as the spin being at maximum $S=N/2$. As discussed above,
for $N=4,~8,~12,~\cdots$ the ferromagnetic state has
a spin-wave (or domains) and interpretation of the
magnetic structure is more difficult. Nevertheless,
results computed for those particle numbers seem to agree
with the phase diagram shown in Fig. \ref{phase1}.
The results in Fig. \ref{phase1} are computed for 
$U=10$. We repeated some of the points for larger values of $U$
and found the same magnetic states.

It is interesting to compare the above results of the generalised
Hubbard model with those of the density functional mean field 
theory\cite{karkkainen2007,karkkainen2007a,karkkainen2007b}.
The qualitative agreement is perfect: In the case of
two $p$-particles per site ($N/L=2$) the system
shows antiferromagnetism of spin-one quasiparticles, while
in the case of one $p$ particle per site ($N/L=1$) the system is ferromagnetic.
An even simpler tight-binding model\cite{koskinen2003} also
gives a similar phase diagram. Due to the symmetry of the Hamiltonian,
it is natural that also above the filling $N/L=2$
one obtains a ferromagnetic region with its center at $N/L=3$.

For small values of $t_2$ the corresponding single-particle band
becomes very narrow. In this case the ferromagnetism can be 
understood with the Stoner mechnanism\cite{mattis1981}: The Fermi level is
in the region of large density of states and induces a 
ferromagnetic state. In 1D this effect is particularly strong
due to the singularities in the density of states~\cite{karkkainen2007}.

\section{Conclusions}

We studied the magnetism of one-dimensional artificial
lattices made of quasi-two-dimensional potential wells,
for up to four particles per lattice
site, i.e. in the region where the $1p$ level is filled.
We froze the $1s$ particles and considered only the $1p$ states.
Numerical diagonalization of a generalized Hubbard model 
was performed for several particle numbers and filling fractions.
The results were analyzed using the antiferromagnetic Heisenberg model 
and single-particle models.

In the resulting phase diagram, the vertical lattice
is ferromagnetic, except at a singular point with exactly
two $p$-type particles per site. 
For lateral lattices the ground state is 
antiferromagnetic for small fillings and close to half-filling
of the $p$-shell, but ferromagnetic around the region with
one $p$-particle per site. A simple model for the ferromagnetic
region was suggested.

If the particle number is a multiple of four
($N=4,8,~12,\cdots$), the ferromagnetic state has
a spin wave which removes the degeneracy and yields a 
total spin $S=0$.

For polarized fermions the half-filled case
shows ``orbital antiferromagnetism'' where in successive
lattice sites the particles rotate clockwise and counter-clockwise.

{\bf Acknowledgments}

This work was supported by the Academy of Finland and by the 
Jenny and Antti Wihuri Foundation, as well as NordForsk, the Swedish Research
Council and the Swedish Foundation for Strategic Research.

\end{document}